\documentclass{article}
\usepackage[utf8]{inputenc}

\title{Neural Network Degeneration and its Relationship to the Brain}
\author{Jacob H. Adamczyk}
\date{\today}

\usepackage{graphicx}
\usepackage{tikz}
\usepackage{neuralnetwork}
\usepackage[square,sort,comma,numbers]{natbib}

\usetikzlibrary{matrix,chains,positioning,decorations.pathreplacing,arrows}
\begin{document}

\maketitle

\begin{abstract}
    This report discusses the application of neural networks (NNs) as small segments of the brain. The networks representing the biological connectome are altered both spatially and temporally. The degradation techniques applied here are ``weight degradation", ``weight scrambling", and variable activation function. These methods aim to shine light on the study of neurodegenerative diseases such as Alzheimer’s, Huntington’s and Parkinson’s disease as well as strokes and brain tumors disrupting the flow of information in the brain's network. Fundamental insights to memory loss and generalized learning dysfunction are gained by monitoring the network’s error function during network degradation. The biological significance of each facet is also discussed.
\end{abstract}

\section{Introduction}
\begin{figure}

    \begin{tikzpicture}[
    init/.style={
      draw,
      circle,
      inner sep=2pt,
      font=\Huge,
      join = by -latex
    },
    squa/.style={
      draw,
      inner sep=2pt,
      font=\Large,
      join = by -latex
    },
    start chain=2,node distance=13mm
    ]
    \node[on chain=2] 
      (x2) {$x_2$};
    \node[on chain=2,join=by o-latex] 
      {$w_2$};
    \node[on chain=2,init] (sigma) 
      {$\displaystyle\Sigma$};
    \node[on chain=2,squa,label=above:{\parbox{2cm}{\centering Sigmoid \\ activation}}]   
      {$f$};
    \node[on chain=2,label=above:Output,join=by -latex] 
      {$y(h_{i})$};
    \begin{scope}[start chain=1]
    \node[on chain=1] at (0,1.5cm) 
      (x1) {$x_1$};
    \node[on chain=1,join=by o-latex] 
      (w1) {$w_1$};
    \end{scope}
    \begin{scope}[start chain=3]
    \node[on chain=3] at (0,-1.5cm) 
      (x3) {$x_3$};
    \node[on chain=3,label=below:Weights,join=by o-latex] 
      (w3) {$w_3$};
    \end{scope}
    \node[label=above:\parbox{2cm}{\centering Bias \\ $b=0$}] at (sigma|-w1) (b) {};
    
    \draw[-latex] (w1) -- (sigma);
    \draw[-latex] (w3) -- (sigma);
    \draw[o-latex] (b) -- (sigma);
    
    \draw[decorate,decoration={brace,mirror}] (x1.north west) -- node[left=10pt] {Inputs} (x3.south west);
    \end{tikzpicture}
    \centering
    \caption{Computational Neuron}
    \label{fig:comp_neuron}
    
\end{figure}
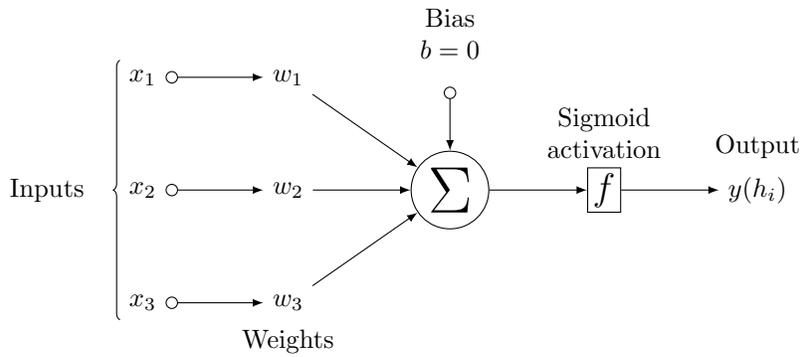

\begin{figure}
    \centering
    \includegraphics[width=0.75\linewidth]{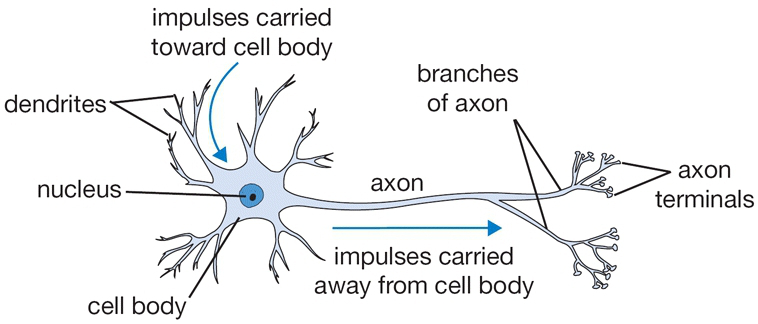}
    \caption{Biological Neuron}
    \label{fig:bio_neuron}
\end{figure}

When initially introduced, artificial neural networks (ANNs) were meant to propagate information in the same way as the dendrite-axon synapse system in the human brain (Figure \ref{fig:comp_neuron}, \ref{fig:bio_neuron}, \ref{fig:my_NN}). This opened the potential for humanistic “learning” in machines. More recently, the original shallow neural network has expanded into a multitude of different varieties, each used for unique tasks. One classic example is the convolutional neural network (CNN) used for image classification. While much has been learned by expanding these networks into increasingly higher abstractions and more complex architectures, not much has been said about the relationship between neurodegenerative diseases (NDs) and NNs. With the successful training of a NN, the questions arise: What happens when the training stage is incoherent? What occurs to a pre-trained NN that is degraded in some way? And how do these questions connect to the biological understanding of the brain? The answers to these questions will be addressed in the following sections.



\begin{figure}

    \begin{neuralnetwork}[height=4]
        \newcommand{\x}[2]{$x_#2$}
        \newcommand{\y}[2]{$\hat{y}$}
        \newcommand{\hfirst}[2]{\small $h_#2$}
        \inputlayer[count=3, bias=false, title=Input\\layer, text=\x]
        \hiddenlayer[count=4, bias=false, title=Hidden\\layer 1, text=\hfirst] \linklayers
        \outputlayer[count=1, title=Output\\layer, text=\y] \linklayers
    \end{neuralnetwork}
    \centering
    \caption{Shallow NN used in this study.}
    \label{fig:my_NN}
\end{figure}

\section{Background}
There currently stands a vast array of NN architectures to study with respect to the brain. The most biologically realistic of which is currently the spike-train NN, which is still largely considered to be a work in progress \cite{SNNref}. Instead, the primary focus of this study is the shallow neural network, with input, output, and a single hidden layer. This network is fully connected on a layer-by-layer basis as depicted in Figure \ref{fig:my_NN}. The purpose of using the simplest use case possible is to observe the most fundamental ideas without a high level of abstraction present in other higher order architectures. In the case of a shallow NN, a binary function can be easily learned by the network. Within 1500 iterations, the error between this neural network's derived function and the desired binary function is typically within 5\%.

The network’s edges are initialized with random weights, which are fine-tuned in the learning process. The learning method used in this network is the standard “feedforward” and “backpropagation” iteration. Feedforward pushes the $X$ data forward with global sigmoid activation functions provides an output vector, \( \tilde{y}\) which is desired to be close to the true function's value, \(f(X) = y\). Closeness between \(\tilde{y}\) and \(y\) simply refers to the RMS difference between the two vectors:

\begin{equation}\label{eqn:error}
    E = \sqrt{\sum_{i} (\tilde{y}_{i}-y_{i})^2 }
\end{equation}

This measurement of the error is the essential quantity studied throughout this report.
Backpropagation is used to work backwards in the network from output to input. By moving right to left, the weights are updated based on the differential change in errors. In order to accurately learn the truth table’s output function $y = f(X)$, the network performs feedforward and backpropagation thousands of times.  These cycles are referred to as \textit{epochs}.
The biological actions of learning and memory recall are very complex activities that are not very well understood. While it will not be possible to fully understand these mechanisms with such a simple NN, this case study aims to reveal the most basic dynamics present in both the computational and biological brains.

\section{Degradation Techniques}
There are interesting spatiotemporal aspects of neural networks which are discussed in this section to motivate the types of simulations involved later on. The network can be altered with respect to time as well as with respect to its structure.

There are two distinct temporal classes of degrading a NN with arbitrary architecture. Those classes are split between degrading the NN
\begin{itemize}
    \item While the network is learning, or
    \item After the network has learned.
\end{itemize}

Due to the fundamental graph-theoretic structure of any NN (Figure \ref{fig:my_NN}), the degradation may also be broken down into two spatial classes. The debilitation can occur either in terms of the associated graph's:

\begin{itemize}
    \item Edges, which carry associated multiplicative weights, or
    \item Vertices, which sum the input and send to the next layer via an activation function.
\end{itemize}

These situations have parallel biological analogies. Network degradation while learning represents the occurrence of a learning impairment in which the patient’s ability to learn and memorize is hindered by some mechanism in the brain. 
Network degradation on a pre-trained model represents the reduction in the patient’s ability to recollect certain ideas due to physical impairment or after-the-fact damage to the brain cells.

\subsection{Degradation on Edges}
\subsubsection{Weight Scrambling}
One approach to neural degradation is to reduce the effectiveness of certain, typically fine-tuned, weights in the NN by ``scrambling” them. To do this, a random number is chosen from a normal distribution centered at 1 between two floats \(1 \pm \sigma\), where \(\sigma\) is henceforth referred to as the ``scrambling variance". The particular value of the scrambling variance in each simulation turns out to be of great importance, and may be the deciding factor in the very existence of a memory-related illness or learning disorder.

\subsubsection{Neuronal Weakening}
Most neurological deficiencies are defined by the general death of neurons or zones in the brain \cite{synapoptosis}. One interpretation of this death is the network's drift from optimization via scrambling (see above). An alternative approach to the degradation of the network is to attack the fine tuned weights by successively multiplying them by some value, \(0 < D < 1\). Over time, the weights will reduce to zero and not lead to activated output for the proceeding layer. Simply deleting the weight ($D=0$) is non-physical and would not allow for studying the intermediate death dynamics.

\subsection{Degradation on Vertices}
The vertices, or nodes, in a NN may be degraded by altering the activation function - the role of the vertex, or axon. Deviation from the standard activation function is another way of shifting the network away from its equilibrium state: minimized error. The results of changing the learning scheme by weakening or strengthening the activation functions are shown in a later section. 

\section{Results}
\subsection{Dysfunctional Learning}
\begin{figure}[h!]
     \centering
     \includegraphics[width=0.75\linewidth]{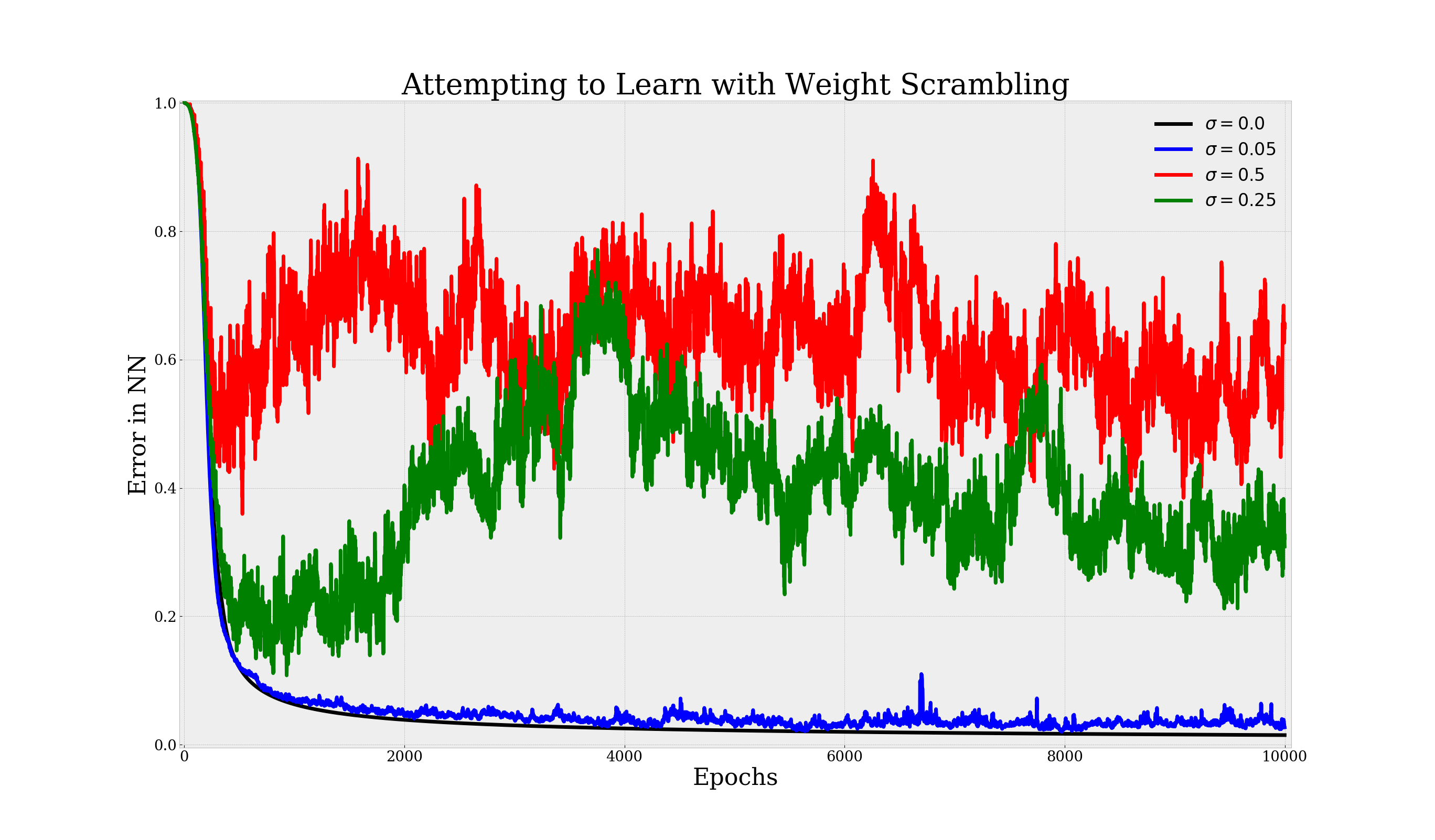}
     \caption{Depending on $\sigma$, effectiveness of learning can be greatly obstructed.}
     \label{fig:WS1}
 \end{figure}

Studying a learning disorder in this framework amounts to applying a neural degradation technique while the NN is training. The network has been trained with and without (as a control) scrambling over 10000 epochs. This experiment is done with different scrambling variances. The Error function (Equation \ref{eqn:error}) is plotted as a function of the epoch in Figure \ref{fig:WS1} for several values of $\sigma$. Each variance experiment is performed 50 times in order to average over some of the statistical noise.
 
During the first several hundred epochs, scrambling is not a significant factor in the error. However, around 500 epochs, the error functions begin to noticeably vary. For a variance \(\sigma \approx 5.0\%\) there is not much effect when comparing the standard error function, with no forced weight scrambling. This is because the randomization of the weights can help move the weights in their optimization landscape towards a minimum. This helpfulness of the randomness is only valid to some extent, because beyond \(\sigma \geq 5.0\%\) weight scrambling, long-term training reveals that the error may be bounded below, and the network is not likely to train well enough to substantially reduce the error. Reducing the error to below 0.5\% is simple enough for no- or low-scrambling networks since longer training times continue to reduce the error. However, at 50\% variance (i.e., multiplying a randomly chosen weight by a random number between 0.5 and 1.5) during each training interval, there is a persisting error around 60\%.

Therefore, no matter the length of time a patient has to study and memorize a particular object (most applicably, a list of numbers or graphical representation of the binary function), a significant non-zero error will remain.
This way of modelling brain illnesses or diseases is such that the weights connecting neighboring nodes are scrambled. From the biological side, this effect is characterized by a defectiveness in the brain’s optimization strategy. This defectiveness can evolve over time and is here conjectured to be the result of:
\begin{itemize}
    \item Weakness presence in the axon terminals or dendrites, debilitating their communication 
    \item The presence of electrochemical noise between neighboring synapses
\end{itemize}
These conjectures are supported by the experimental findings of Andreotti et. al. \cite{Explanation}: ``The analyses showed that Alzheimer's disease is characterized by decreased connectivity strength in various cortical regions." 

The exact relationship between scrambling variance and the long-term persisting error is shown below. The error is averaged over 50 iterations and from 500 to 10000 epochs to reduce the impact of statistical fluctuations and large error terms in the beginning of the NN’s lifetime, respectively. The standard deviation among the 50 iterations as well as the average is plotted below. 

\begin{figure}[h!]
    \centering
    \includegraphics[width=0.75\linewidth]{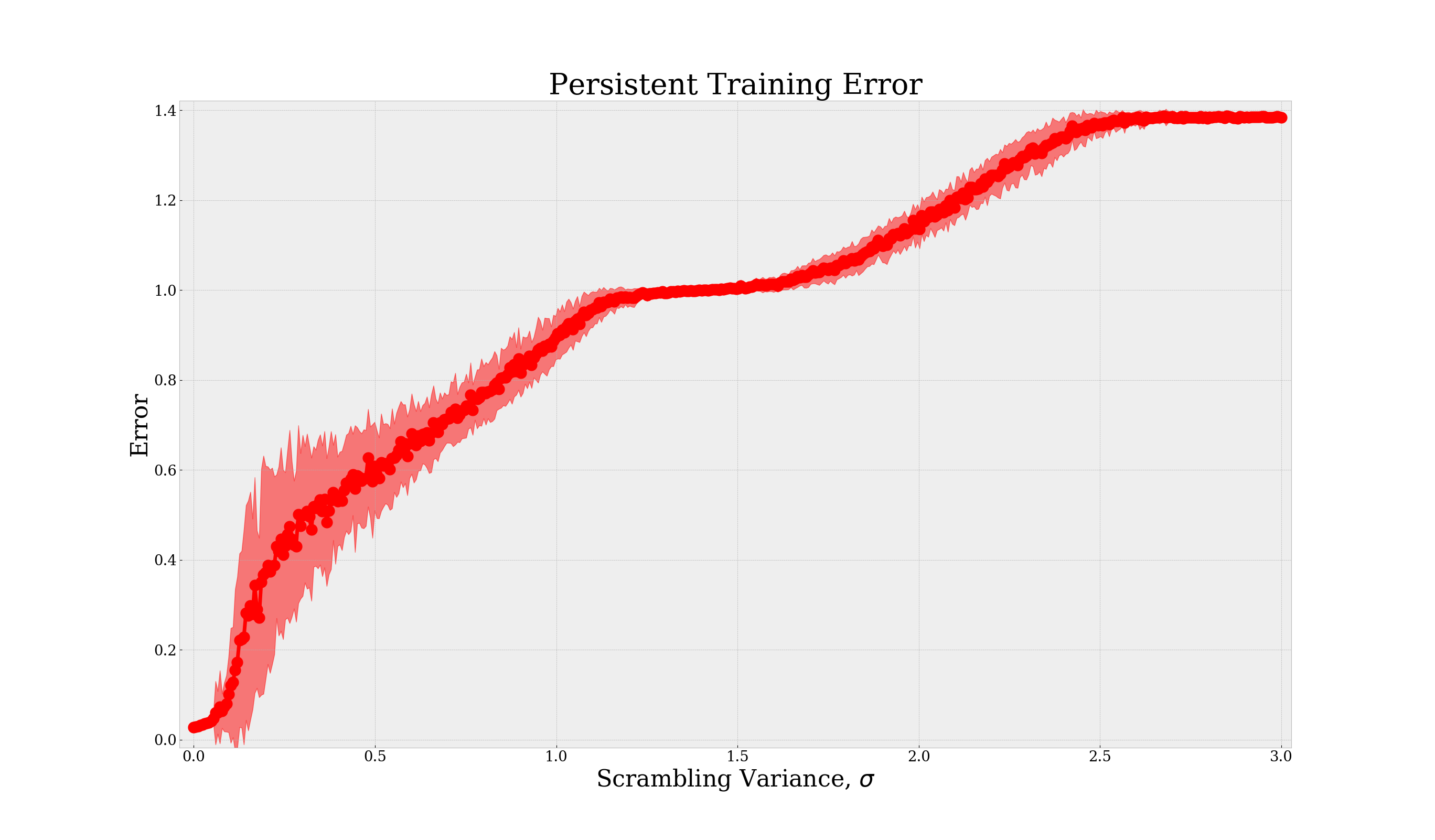}
    \caption{Residual error, for different values of $\sigma$.}
    \label{fig:my_label}
\end{figure}

\underline{Medical Context:}
Patient A is able to learn or memorize the object in question (no- or low-scrambling environment), and Patient B is able to learn or memorize with a measured error of \(E\). The error, \(E\) can be calculated from the trial’s results. This value of \(E\) can be matched with the plot above to determine the internal scrambling \(\sigma\) in Patient B’s biological network. The above simulations suggest that an experiment and outcome of this type may be the result of widespread network defectiveness caused by improper synaptic communication. The calculated value of \(\sigma\) can be used as a quantitative measure in the diagnosis of Patient B. If a drug existed to treat this issue and flush out the excess tau protein aggregation \cite{Explanation}, for example, the calculated value of \(\sigma\) could hypothetically dictate the required dosage of such a treatment.

\subsubsection{Reduced and Heightened Activation Sensitivity}

With weaker or fewer axonal cells, the activation function, which determines how important particular values are, and how strongly they are activated in the next layer, may be reduced in its sensitivity. This is depicted in the 50\% reduced sigmoid. Incoming signals must be larger in magnitude to provide a proper activation relative to the standard sigmoid. The opposite may be said of the heightened sigmoid, where input data is more easily activated due to higher sensitivity.

\begin{figure}
    \centering
    \includegraphics[width=0.75\linewidth]{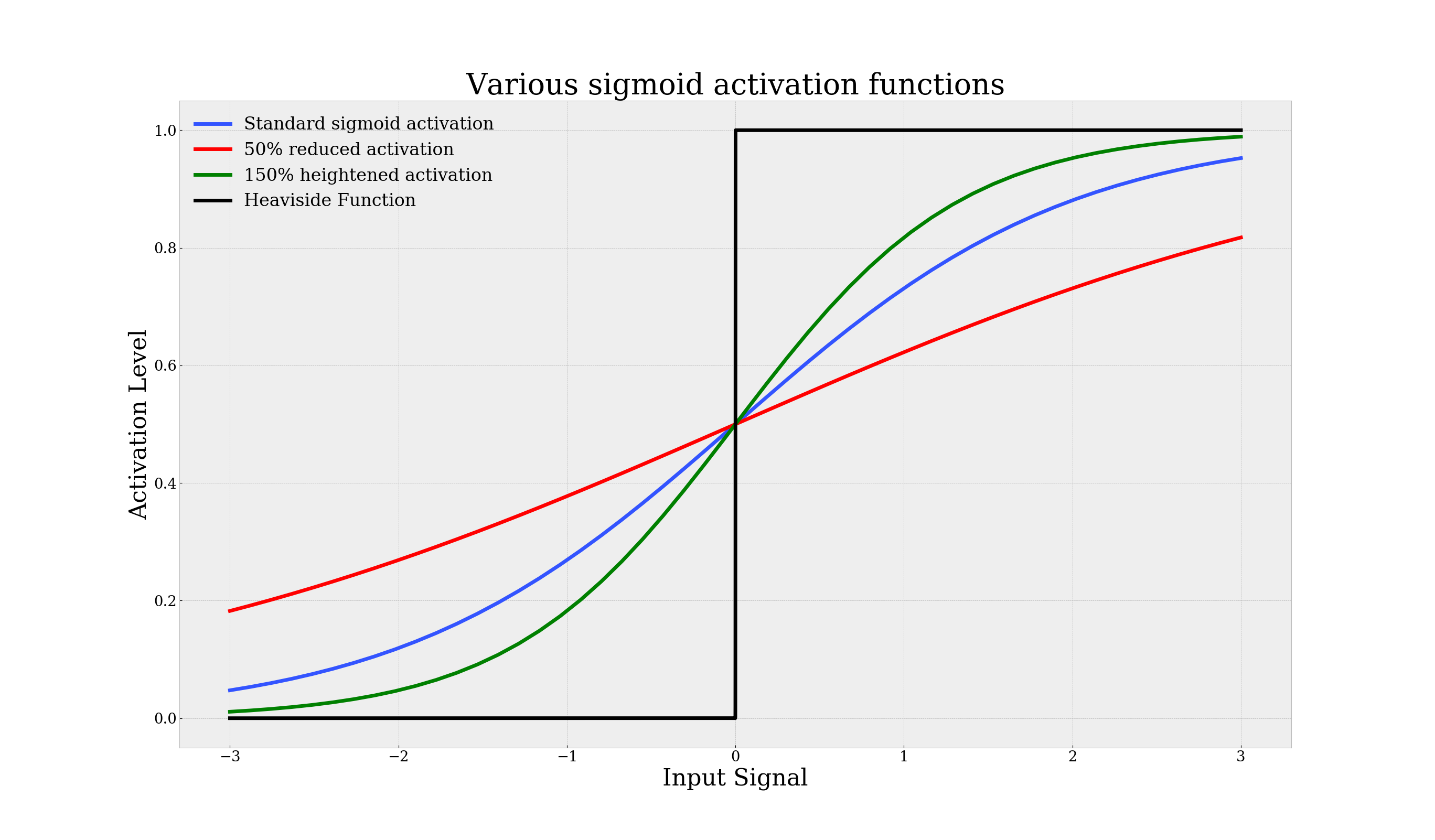}
    \caption{Different Sigmoid Activation Functions}
    \label{fig:sigmoid}
\end{figure}

Below, the results of learning with different sensitivities shows that a 50\% reduced sigmoid marks a significant increase in the amount of time required to reduce the error to the same amount as in the case of the standard sigmoid. To reach an error below 5\%, the high sensitivity network requires 450 epochs of training; for the standard sensitivity network 610 epochs, and for the low-sensitivity, 890 epochs. However, beyond a 200\% heightened activation strength, the sigmoid activation begins to resemble a binary Heaviside function, as shown in Figure \ref{fig:sigmoid}. The Heaviside function is piecewise, and clearly is non-differentiable at zero and its derivative is zero elsewhere. Due to the nature of this function's derivatives, the gradient descent algorithm for backpropagation does not always work well. Hence, this extremely heightened type of activation does not prove to be useful for NNs and this is the reason why sigmoid functions replaced them historically.

 \begin{figure}
     \centering
     \includegraphics[width=0.75\linewidth]{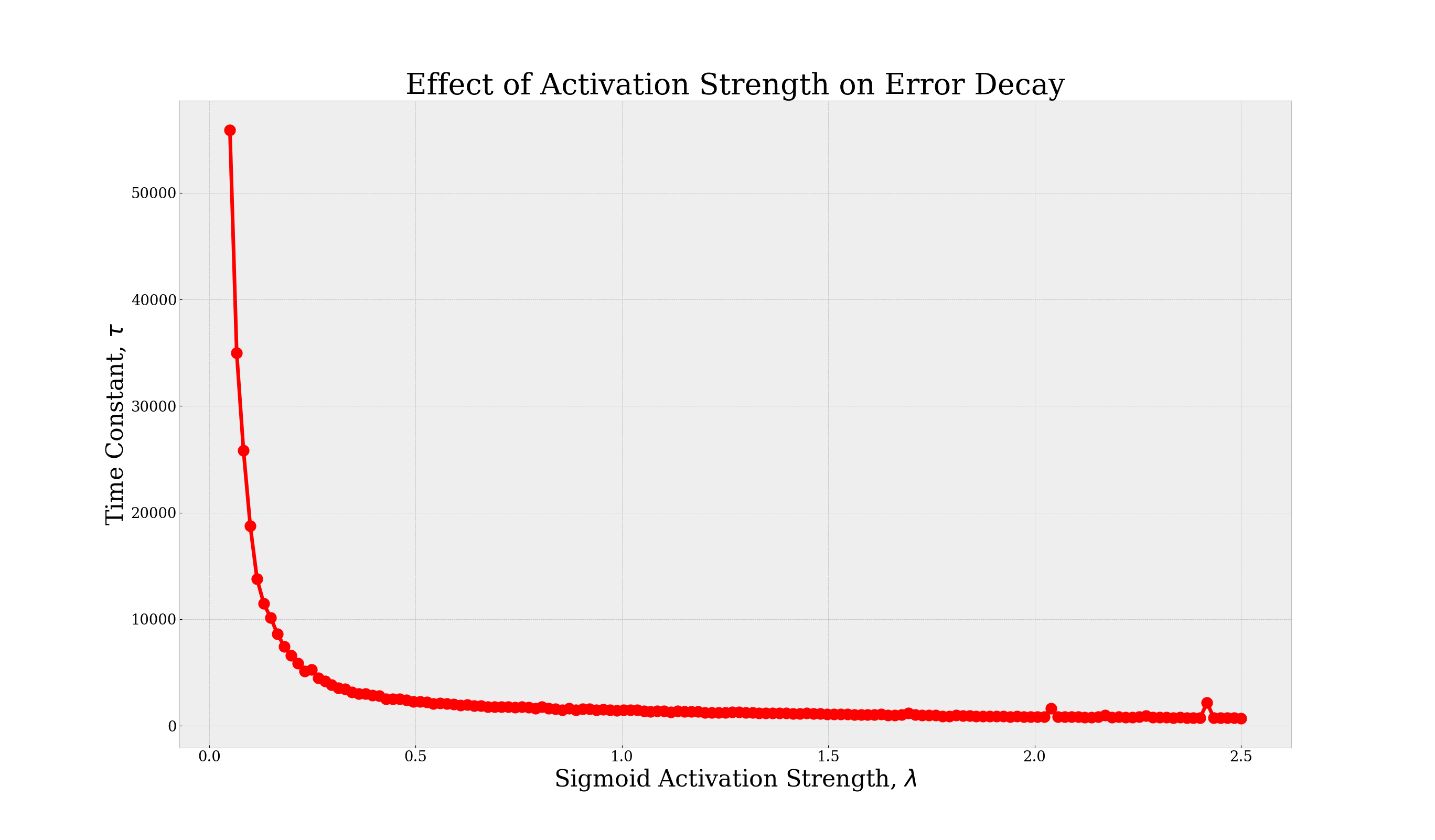}
     \caption{The time to reach 5\% error reduces greatly for incrementally more heightened activation functions}
     \label{fig:diff_sigmoid_results}
 \end{figure}

\underline{Medical Context:}
Compare two patients: Patient A has standard, healthy activation functions, with $\lambda \approx 1.0$. Patient B has weakened neuron activation functions. Trials between the two show that Patient A (on average) takes time $T_A$ to memorize the object of study. This measurement can be used to normalize the conversion between network epochs and physical time steps. Measurements also show Patient B (on average) takes time $T_B$ to memorize the same object. From this clinical study, one may calculate the magnitude of a widespread (averaged) weakness within Patient B’s neuronal activation functions. This method may be used to quantitatively differentiate between activation strengths among patients, and indicate progression or severity of a learning disorder.

\subsection{Memory-related disabilities}
In the previous section, the effect of scrambling during training was studied. This section is concerned with the second approach outlined above: to train a network fully, and \textit{then} scramble its weights, while observing the effect on the error. The biological significance here is that a fully developed brain subsystem (NN) is well-trained and later in time, there is a degradation in the brain causing the already learned information to be lost to the problematic brain cells. Again, the discussion is in terms of scrambling the weight factors of random edges in the NN.

\subsubsection{Weight Scrambling}

In this experiment, the weight scrambling variance \(\sigma\) ranges from 0.05 to 1.00. The amount of iterations in scrambling physically corresponds to a time constant with arbitrary units but presumably constant conversion factor. Whatever the conversion may be between these iterations into units such as minutes or years, it is assumed to be a constant multiplicative factor. The 50\% error cutoff is the characteristic timescale at which most meaningful knowledge in the network is lost. Therefore, the NN is degraded until its error surpasses 50\% and this time (epoch), $\tau$ is recorded.
  \begin{figure}
     \centering
     \includegraphics[width=0.75\linewidth]{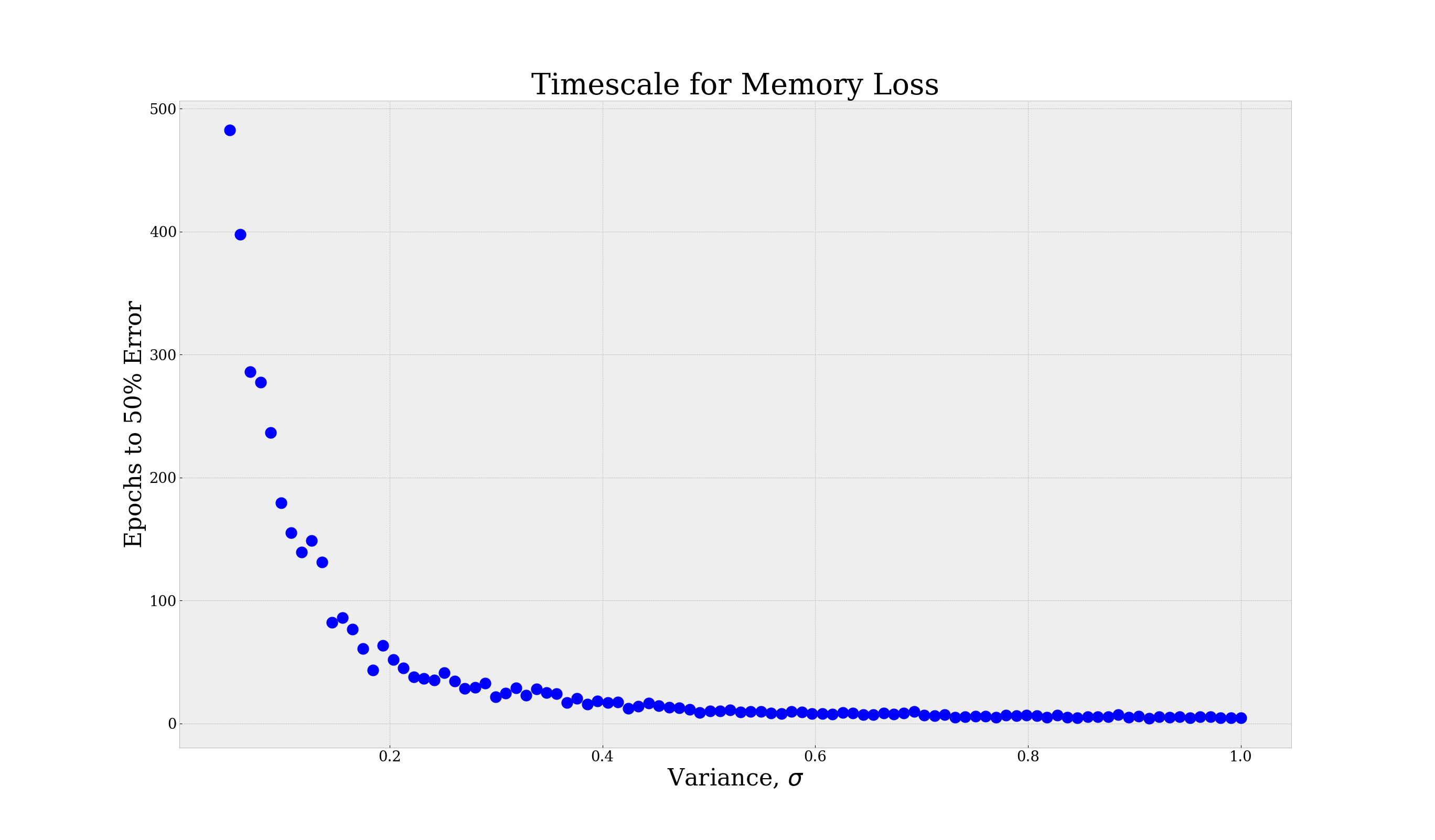}
     \caption{Characteristic timescales for different values of \(\sigma\).}
     \label{fig:mem_loss}
 \end{figure}

\underline{Medical Context:} Consider Patient A and Patient B being tested on the same arbitrary object. If it is found that the scrambling within Patient A is $\sigma_A$, then this information gives an estimate $f(\sigma_A)$ on the amount of time until Patient A has a 50\% error on the arbitrary object and no longer holds any useful information in their own NN. If a ND is assumed to worsen in Patient B over time \cite{Khan2016}, one may view the above horizontal axis as duration of disease within Patient B. While there is no cure for many NDs, it is possible in some cases to slow down or stop its effects \cite{slowNDrats} \cite{slowingHandP}. If the disease is diagnosed and treated, $f(\sigma_B)$ provides an upper bound to the amount of time until full degradation will occur. Assuming the former, an early diagnosis ($\sigma = 0.10$) can yield a time constant of 200 (say, days) until full degradation; whereas a slightly later diagnosis ($\sigma = 0.3$) yields a time constant of only 30 days until the same amount of memory is lost. 

Notice that this does not account for a possible non-linear relationship between scrambling variance and duration of disease presence. If there is a functional relationship between the two then a better model may be developed by converting between the two. If \(\sigma\) is interpreted as a metric for the severity of the ND, Khan has suggested that for normal aging processes, the functional form of $\sigma(T)$ is linear\cite{Khan2016}. If, instead, symptoms of AD are traceable, then a transition from normal ageing neural degeneration occurs to an increased rate, or slope, in $\sigma(T)$. This implies an acceleration of the already steep decrease in memory loss shown in Figure \ref{fig:mem_loss}.

\subsubsection{Decaying Neurons}
Beginning with the fine-tuned neuronal weights $w_i$, a decay constant $D \in (0, 1)$ is successively multiplied at each epoch: $w_i(n) = w_i D^n$. An increase in the decay constant, $D$ will lead to a slower speed of degradation for the network since multiplying by values close to unity do not have as significant effect on the NN compared to smaller values. A slower speed of degradation ($D \xrightarrow  \ 0$) would imply a larger time constant $\tau$, implying a longer lasting memory in the NN. An asymptotic behavior is also expected such that in the limit $D \xrightarrow  \ 0$, the time constant $\tau \xrightarrow \ \infty$. If there is no decay in the network, the error will never surpass 50\%. This explanation matches the simulation's result shown below. 

\begin{figure}[h!]
    \centering
    \includegraphics[width=0.75\linewidth]{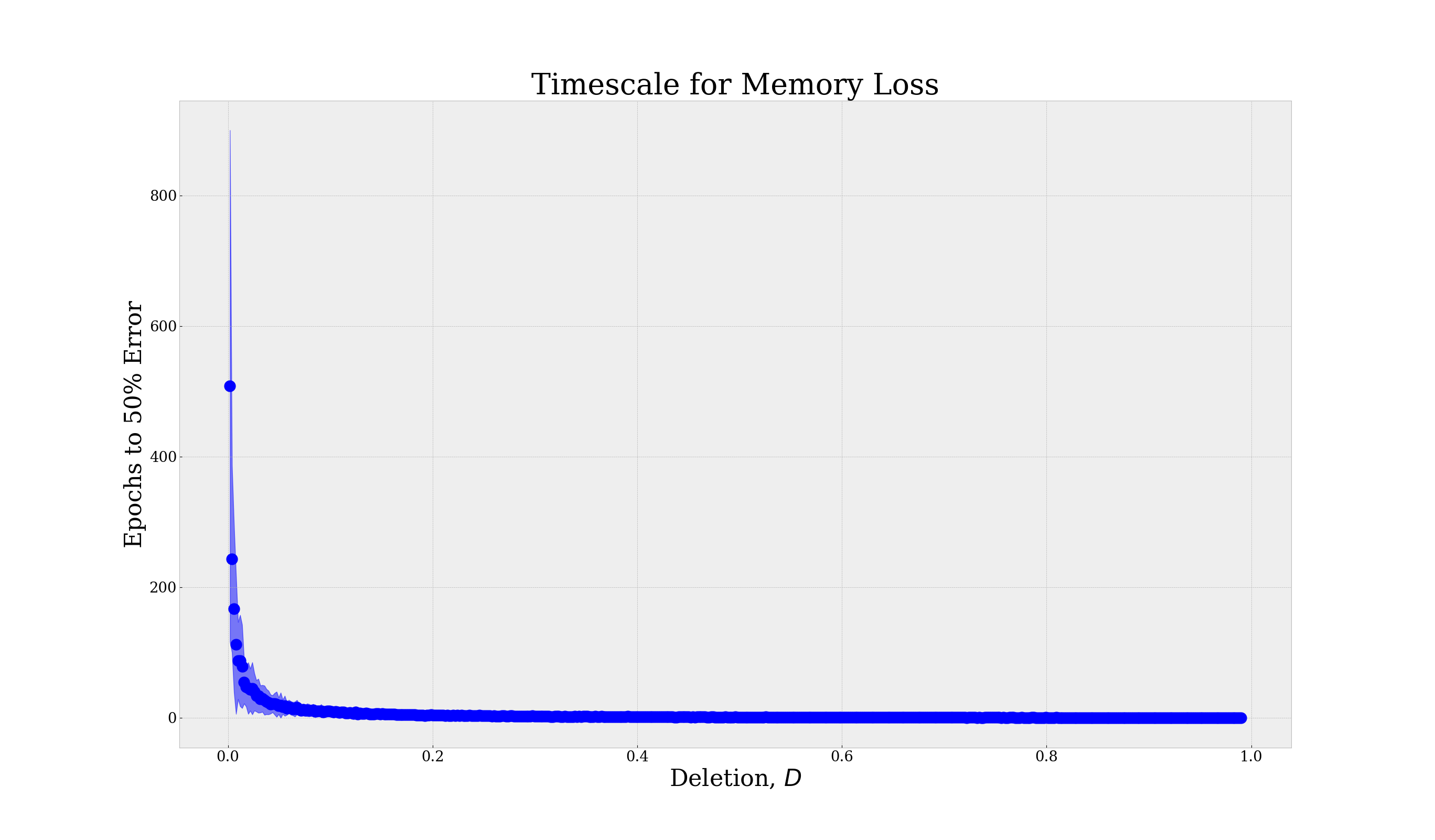}
    \caption{Results of degrading randomly chosen weights.}
    \label{fig:delet}
\end{figure}

In this framework, if converting from true time to network epochs is possible, a calculated value of the deletion parameter provides more information about the progression or severity of the ND. 

\section{Conclusions}
Clinical experiments may be designed to calculate the scrambling variance ($\sigma$), activation strength ($\lambda$), or decay constant ($D$), which could then provide useful information regarding diagnosis and dosing.

Computational experiments regarding learning and memory-related disorders have been examined with a shallow neural net as the driving model. 
Using neural networks to study the brain can be an important avenue of research, especially in regard to neurodegenerative diseases and learning disabilities. These may be studied by deterioration of the network's parameters during or after training. This report highlights the fundamental tools for these ideas and explores possible paths for connecting the methods of biology and computer science.

\section{Future Work}
Future research in this direction involves a change of scope to spike-based neural networks \cite{SNNref}. More complicated and hypothetically more realistic NN models such as deep and long short-term memory may provide useful case studies for this avenue as well. 

\bibliographystyle{plain}
\bibliography{references}
\end{document}